\documentclass[preprint,12pt,3p]{elsarticle}

\usepackage{graphics}
\usepackage{graphicx}
\usepackage{epsfig}

\usepackage{amssymb}
\usepackage{xcolor}
\usepackage{times}
\usepackage{epsfig}
\usepackage{graphicx}
\usepackage{subfig}
\usepackage{float}
\usepackage{amsmath}
\usepackage{amssymb}
\usepackage{algorithm}
\usepackage{algorithmic}
\usepackage{multirow}
\usepackage{setspace}

\makeatletter \@fpsep = 2pt \makeatother
\journal{CAAI Transactions on Intelligence Technology}

\begin{document}

\begin{spacing}{1.0}

\begin{frontmatter}

\title{Image super-resolution via dynamic network}

\author[label1,label2,label3]{Chunwei Tian\corref{cor1}}
\author[label1]{Xuanyu Zhang}
\author[label4]{Qi Zhang*}
\ead{hit_zq910057@163.com}
\author[label5]{Mingming Yang}
\author[label6]{Zhaojie Ju}

\address[label1]{School of Software, Northwestern Polytechnical University, Xi’an, 710129, China}
\address[label2]{National Engineering Laboratory for Integrated Aero-Space-Ground-Ocean Big Data Application Technology, Xi’an, 710129, China}
\address[label3]{Research \& Development Institute, Northwestern Polytechnical University, Shenzhen, China}
\address[label4]{School of Economics and Management, Harbin Institute of Technology at Weihai, Weihai, 264209, China}
\address[label5]{Tencent AI lab, Shenzhen, China}
\address[label6]{School of Computing, University of Portsmouth, Portsmouth PO1 3HE, UK}

\begin{abstract}
Convolutional neural networks depend on deep network architectures to extract accurate information for image super-resolution. However, obtained information of these convolutional neural networks cannot completely express predicted high-quality images for complex scenes. In this paper, we present a dynamic network for image super-resolution (DSRNet), which contains a residual enhancement block, wide enhancement block, feature refinement block and construction block. The residual enhancement block is composed of a residual enhanced architecture to facilitate hierarchical features for image super-resolution. To enhance robustness of obtained super-resolution model for complex scenes, a wide enhancement block achieves a dynamic architecture to learn more robust information to enhance applicability of an obtained super-resolution model for varying scenes. To prevent interference of components in a wide enhancement block, a refinement block utilizes a stacked architecture to accurately learn obtained features. Also, a residual learning operation is embedded in the refinement block to prevent long-term dependency problem. Finally, a construction block is responsible for reconstructing high-quality images. Designed heterogeneous architecture can not only facilitate richer structural information, but also be lightweight, which is suitable for mobile digital devices. Experimental results show that our method is more competitive in terms of performance, recovering time of image super-resolution and complexity. The code of DSRNet can be obtained at https://github.com/hellloxiaotian/DSRNet.
\end{abstract}


\end{frontmatter}


\section{Introduction}
\label{sec-1}

Single image super-resolution (SISR) is a basic task of image restoration \citep{wang2020deep}. It can use an image degradation model to convert a low-resolution (LR) image to recover a high-resolution (HR) image \citep{yang2019deep}. Also, some interpolation methods, i.e., nearest neighbor interpolation \citep{rukundo2012nearest}, bilinear interpolation \citep{li2001new} and bicubic interpolation \citep{keys1981cubic} were used to achieve image degradation models for image super-resolution. Although they were simple, they had poor performance in image super-resolution \citep{park2003super}. To overcome this challenge, optimization methods were developed to recover high-quality images \citep{yang2022non, ZhaA2020, ZhaFrom2020}. For instance, an iterative back-projection (IBP) \citep{yoo2019noise} can iteratively calculate a distance between simulated LR and observation LR images to continuously update predicted HR images. When calculated distance is less than a given threshold, predicted HR images are final high-quality images. Although this method can obtain good effects on image super-resolution, it referred to complex optimization methods, which may increase computational cost. Also, it also referred to manual setting parameters. 
  
To address this problem, deep learning techniques are proposed for image applications, i.e., image classification\cite {cmc.2023.040560} and image super-resolution \cite{wang2023lightweight}. For instance, Wang et al. \cite {wang2023pstcnn} used particle swarm optimisation to guide a convolutional neural network (CNN) to automatically adjust parameters to improve recognition efficiency of COVID-19. Wang et al. \cite{wang2022covid} combined wavelet entropy and a neural network to better represent images to improve recognition rate of COVID-19 diagnosis. Also, deep learning techniques have obtained good effects on food recognition \cite{zhang2023deep}. Also, CNNs are strong self-learning abilities. Thus, CNNs have been extended in the field of image super-resolution. For instance, Dong et al. used a 3-layer network architecture in a pixel mapping manner to achieve a mapping from a LR image to a HR image \citep{dong2014learning}. Although it has obtained better performance than that of traditional methods for image super-resolution, it has poor scalability. To solve this problem, Kim et al. used a residual learning operation to act a deep layer in a deep network to extract accurate information to promote visual effect of image super-resolution \citep{kim2016accurate}. To avoid gradient vanishing and exploding phenomenon, Kim et al. gathered a recursive supervision and skip connection to enhance learning ability of a designed network in image super-resolution \citep{kim2016deeply}. Alternatively, Yang et al. used gate unit, skip connections and dense connection architecture to inherit obtained features of shallow layers to deep layers for enhancing memory abilities of shallow layers for image super-resolution \citep{tai2017memnet}. Although these methods perform well for image super-resolution, their inputs are bigger, which may cause big computational costs. To resolve this issue, Dong et al. directly used low-resolution images as an input of a deep CNN and an up-sampling operation in a deep layer to amplify obtained low-frequency features for recovering a high-quality image \citep{dong2016accelerating}. Although it may have little computational costs, it neglected effects of high-frequency features to obtain poor results of image super-resolution. To tackle this problem, a residual dense network used hierarchical information of a deeper network to enhance obtained low-frequency features to capture more detailed information for image super-resolution \citep{hui2018fast}. Tian et al. \citep{tian2021asymmetric} used a symmetrical architecture to enhance low-frequency information for image super-resolution. Although these methods are effective for image super-resolution, they depend on deeper architectures to achieve a good performance, which may result in big computational costs. Besides, obtained information of these CNNs cannot completely express predicted high-quality images for complex scenes.

In this paper, we present a lightweight dynamic network in image super-resolution. It uses a residual enhancement block, wide enhancement block, feature refinement block and construction block to achieve a good result for image super-resolution. The residual enhancement block utilizes a residual learning operation to act a stacked architecture to enhance effects of hierarchical layers. To enhance robustness of an obtained super-resolution for complex scenes, a wide enhancement block enlarges receptive field via combining a stacked architecture and a selected dynamic gate block to extract more accurate information for enhancing applicability of obtained super-resolution model for varying application scenes. To prevent interference of components in a wide enhancement block, a refinement block utilizes a stacked architecture to accurately learn obtained features. Also, a residual learning operation is embedded in the refinement block to prevent long-term dependency problem. Finally, a reconstruction block is responsible for constructing high-quality images. The proposed super-resolution model is very superior to performance, recovering time, and complexity for image super-resolution.

This paper has the following contributions. 

(1) Dynamic gate is used for image super-resolution, which can improve robustness of obtained super-resolution model for varying scenes. 

(2) A progressive feature enhancement and refinement method is proposed to make a tradeoff between performance and complexity of image super-resolution.

(3) The proposed network is lightweight, which is very suitable on real digital devices for image super-resolution.
   
Remaining parts of this paper can be summarized as follows. Section 2 provides a review of related work about lightweight CNNs and dynamic neural networks for image super-resolution. Section 3 provides a detailed introduction of our proposed method. Section 4 lists method analysis and experimental results. Section 5 gives a summary of this paper. 

\section{Related work}
\subsection{Lightweight convolutional neural networks for image super-resolution}
Due to excellent expressive abilities, CNNs have become effective tools to extract accurate information for some applications, music \citep{wang2022cps} and videos \citep{yuan2021active}, especially image processing applications. However, some CNNs reply on increasing depth of networks to improve performance of image processing applications, which will result in big computational costs. Also, it is not suitable to real digital devices. To address this issue, lightweight CNNs are conducted in image super-resolution \citep{hui2019lightweight}. For instance, Hui et al. \citep{hui2019lightweight} utilized a distillation module to obtain hierarchical information and aggregated features to reduce computational costs for image super-resolution, according to their importance. Liu et al. \citep{liu2020residual} used a residual block to improve a deep network from Ref. \citep{hui2019lightweight} to make a tradeoff between performance and computational costs for image super-resolution. Alternatively, Luo et al. designed two butterfly architectures via combining attention mechanisms and residual learning architectures to achieve a lightweight network in image super-resolution \citep{luo2020latticenet}. To deploy on mobile phones, Li et al. \citep{li2020s} exploited a U-Net and depthwise separable convolutions to reduce parameters to improve execution speed for image super-resolution. To recover more texture information, Zuo et al. \citep{zou2022self} designed a serial lightweight architecture via using a self-calibrated module, pixel attention mechanism and Transformer to capture context information for restoring high-quality images. Also, this idea is extended to remote-sensing image restoration. For instance, Wang et al. \citep{wang2021contextual} exploited a contexture transformer layer to mine more detailed information for remote-sensing image super-resolution. Taking into difference of different architectures account, Gao et al. \citep{gao2023very} combined multiple blocks, including a multi-way attention block, lightweight convolutional block to implement a lightweight CNN to facilitate more accurate information for improving effects of image super-resolution, according to different views. Besides, using fast Fourier Transform into a generative adversarial network can enlarge receptive field to improve recovering effects when the number of learned parameters is not increased \citep{nguyen2023f2srgan}. Evolutionary computation and reinforcement learning are used to guide a neural network to obtain a lightweight a network in image super-resolution \citep{chu2021fast}. According to mentioned illustrations, we can see that optimizing a network architecture is effective for achieving a lightweight CNN for image super-resolution. Taking into robustness of obtained information for complex scenes account, we use a dynamic gate, multi-view idea and design principle of a deep network to achieve a lightweight CNN for image super-resolution.

\subsection{Dynamic neural networks for image super-resolution}
It is known that common CNNs are very superior to extracting effective features to improve performance of image applications. However, they cannot flexibly vary as different scenes, which affects their performance in complex scenes. To address this problem, dynamic neural networks are developed, which can dynamically adjust parameters to enhance robustness and stability of obtained models \citep{han2021dynamic}. Due to its superiority, it has been applied for image super-resolution. For example, Shi et al. \citep{shi2020ddet} used a content-adaptive technique containing a multi-scale dynamic attention and a skip connection into a dual network to align images in the real world for image super-resolution. Alternatively, dynamic convolutions are used to adapt varying of between images, between pixels in an image to obtain a flexible super-resolution model for image super-resolution \citep{xu2020unified}. Taking into load capacities of hardware account, discrete cosine transform is used into a network to evaluate complexity of different areas of an image to assign suitable networks for quickly recovering these areas \citep{xie2021learning}. Alternatively, Chen et al. \citep{chen2022arm} assign different networks via edge information to recover detailed information in image super-resolution. To improve performance of image super-resolution, different attention mechanisms are used to respectively extract local and global features to facilitate more useful detailed information for image super-resolution \citep{tian2023multi}. According to mention illustrations, we can see that dynamic networks are effective for image super-resolution. Inspired by that, we design a dynamic network for image super-resolution in this paper.

\section{The proposed method}
\subsection{Network architecture}
It is known that some CNNs trend to increase network depth to improve performance for image processing. However, that may increase computational costs, which is not beneficial for digital devices. Also, single network architecture is not beneficial to extract richer features for complex scenes. To address this problem, we propose a 18-layer progressive network as well as DSRNet for image super-resolution in Figure 1. That is, it is composed of a 6-layer residual enhancement block, a parallel 4-layer wide enhancement block, a 6-layer feature refinement block and 2-layer reconstruction block. Residual enhancement block is used to extract and enhance hierarchical information. Wide enhancement block uses stacked architectures and a dynamic gate to achieve a parallel architecture to facilitate more richer information for complex scenes. To prevent interference of some components in the wide enhancement block, a feature refinement is used to further learn obtained features from the wide enhancement block. To obtain high-quality images, a reconstruction block is used to construct high-resolution images. To simply mentioned illustrations, Eq. (1) is conducted.
\begin{equation}
\begin{array}{l}
{I_H} = DSRNet({I_L})\\
{\rm{\ \ \ \ }} = {\rm{ }}RB{\rm{(}}FEB{\rm{(}}WEB{\rm{(}}REB{\rm{(}}{I_L}{\rm{))))}}
\end{array},
\end{equation}
where ${I_L}$ denotes a low-resolution image, $DSRNet$ expresses a function of DSRNet, $REB$, $WEB$, $FEB$ and $RB$ stand for functions of REB, WEB, FEB and RB, respectively. ${I_H}$ is symbolled as an output of $DSRNet$, which is also a high-quality image. Besides, a loss function can optimize parameters of DSRNet as shown in Section 3.2.

\begin{figure*}[!htb]
\centering
\begin{tabular}{c}
\includegraphics[scale=0.4]{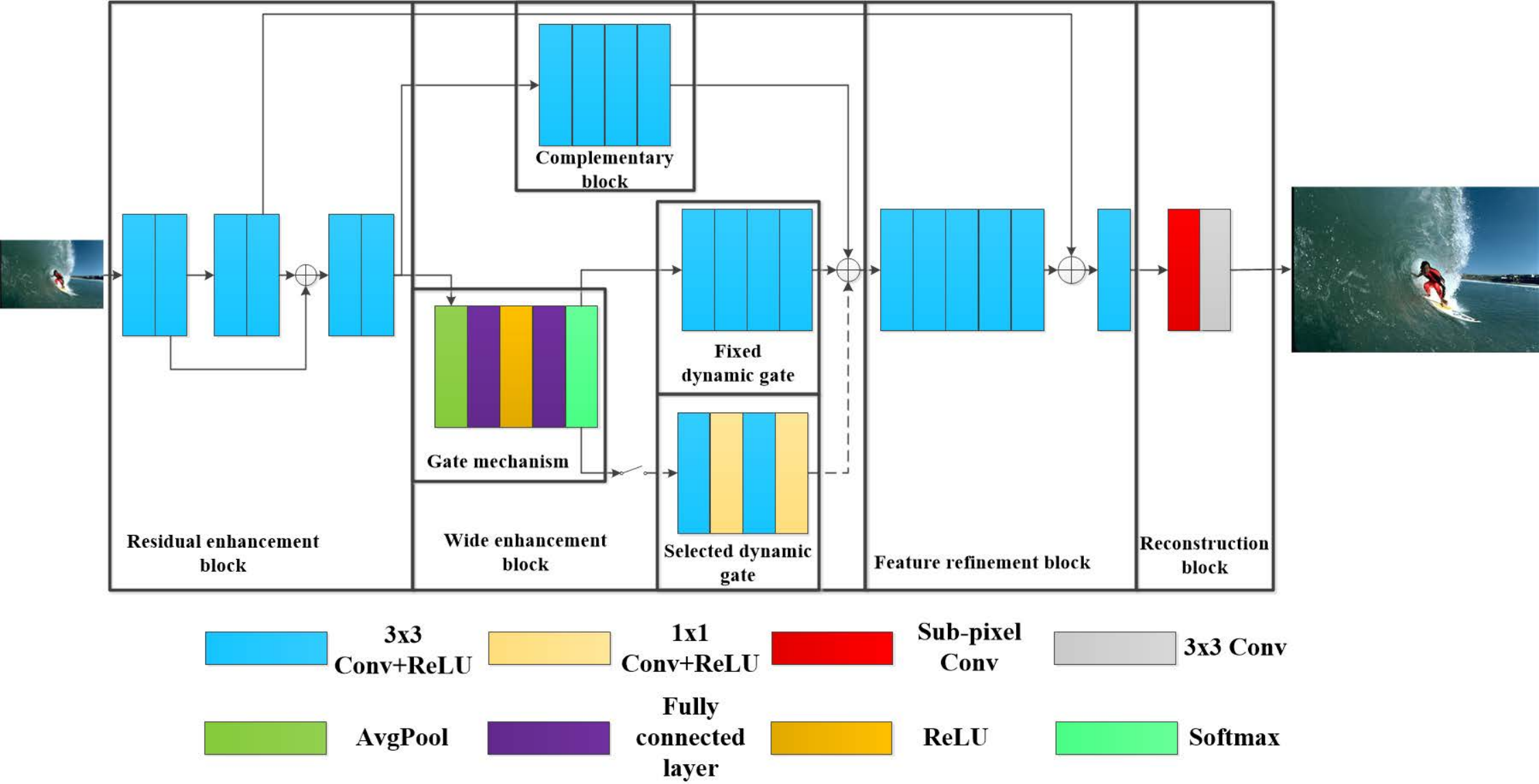}\small
\end{tabular}
\caption{Network architecture of DSRNet.}
\label{fig:1}
\end{figure*}

\subsection{Loss function}
According to Ref. \citep{yang2018drfn, zeyde2012single}, we can see that mean squared error (MSE)\citep{wang2009mean} is a common loss function for image super-resolution. Thus, to fairly obtain a DSRNet, we also choose the MSE as a loss function to train DSRNet model. That is, ${\rm{\{ }}L_L^i{\rm{,}}L_H^i{\rm{\} }}_{i = 1}^n$ from training dataset in Section 4.1 can be acted on the MSE to train a DSRNet model, where $L_L^i$ and $L_H^i$ are respectively regarded as the $ith$ low- and high-resolution training image patches. Let $n$ be the number of training image patches. Descriptions above can be formulated as Eq. (2). 
\begin{equation}
L(p) = \frac{1}{{2n}}\sum\limits_{i = 1}^n {||DSRNet(I_L^i) - I_H^i|{|^2}},
\end{equation}
where $L$ is a loss function of MSE and $p$ can be used to stand for learned parameters.

\subsection{Residual enhancement block}
Residual enhancement block is used to extract and enhance hierarchical information. That is, the residual enhancement block uses six combinations of convolutional layer and ReLU to extract more deep information. Each convolutional layer is used to extract linear features and ReLU is utilized to capture non-linear features. To prevent poor effect of shallow layers on deep layers, we use a residual learning operation to gather outputs of the second and fourth layers to enhance memory ability of shallow layers for image super-resolution. It can be conducted as Eq. (3).
\begin{equation}
\begin{array}{l}
{O_{REB}} = REB({I_L})\\
{\rm{\ \ \ \ \ \ \ \ \ \ }} = 6ReLU(Conv({I_L})) + ReLU(Conv(ReLU(Conv({I_L}))))
\end{array},
\end{equation}
where $Conv$ stands for a function of a convolutional layer, $ReLU$ stands for an activation function of ReLU. + is a residual learning operation, which is as well as $ \oplus $ in Figure 1. $6ReLU(Conv())$ denotes six stacked combinations of a convolutional layer and a ReLU. ${O_{REB}}$ is an output of the residual enhancement block. All the convolutional layers have kernels of $3 \times 3$. Input channels of all the layers besides the first layer are 64. The input channel of the first layer is $c$. And $c$ is same as  channel number of input color images, which is 3. Output channels of all the layers are 64. 

\subsection{Wide enhancement block}
To make more robust information, a 4-layer parallel network architecture as well as a wide enhancement block is designed. That is, a wide enhancement block is composed of two branches. The first branch exploits a 4-layer Conv+ReLU composed of a complementary block to extract more useful deep information, where Conv+ReLU denotes a combination of a convolutional layer and ReLU. To obtain richer information, the second branch is composed of a dynamic network. That is, it uses a gate mechanism composed of an average pooling (AvgPool), a fully connected layer, a ReLU, a fully connected layer and a Softmax to achieve a gate, i.e., selected dynamic gate. Specifically, output of the gate mechanism has two values. The second result is used as a choosen gate. If the second result is lower than 0.75, output of the gate mechanism is inputs of a fixed dynamic gate and a selected dynamic gate. Otherwise, its output is only an input of a fixed dynamic gate. Mentioned fixed dynamic gate contains four Conv+ReLU. The selected dynamic gate includes two stacked combinations of Conv+ReLU with kernel of $3\times3$ and Conv+ReLU with kernel of $1\times1$. To extract complementary features, outputs of a complementary block, a fixed dynamic gate or outputs of a complementary block, a fixed dynamic gate and a selected dynamic gate are gathered via residual learning operations. Specifically, all mentioned convolutional kernel sizes are $3\times3$, input and output channels of all the layers in complementary block, fixed dynamic gate and selected dynamic gate are 64. The process can be expressed as Eq. (4).
\begin{equation}
\begin{array}{l}
{O_{WEB}} = WEB({O_{REB}})\\
{\rm{\ \ \ \ \ \ \ \ \ \ \ \ = (}}CB{\rm{(}}{O_{REB}}{\rm{) + }}FDG{\rm{(}}GM{\rm{(}}{O_{REB}}{\rm{)))}}
\Theta {\rm{SDG(}}GM{\rm{(}}{O_{REB}}{\rm{))}}
\end{array},
\end{equation}
where $CB$, $GM$, $FDG$ and ${SDG}$ denote functions of a complementary block, gate mechanism, fixed dynamic gate and selected dynamic gate, respectively. ${O_{WEB}}$ is an output of the wide enhancement block, which is also an input of feature refinement block. $\Theta $ denotes a selected gate. If the second result of gate mechanism is lower than 0.75, $\Theta $ denotes a residual learning operation. Otherwise, $\Theta $ denotes dropped part. And + is used to represent a residual learning operation, which is as well as $ \oplus $ in Figure 1.

\subsection{Feature refinement block}
To prevent interference of some components in the wide enhancement block, a feature refinement is used to future learn obtained features from the wide enhancement block. It is composed of six stacked Conv+ReLU. To prevent long-term dependency problem, a residual learning operation is acted between outputs of the fourth layer in the residual enhancement block and fifth layer in the feature refinement block.
\begin{equation}
\begin{array}{l}
{O_{FRB}} = FRB({O_{WEB}})\\
{\rm{\ \ \ \ \ \ \ \ \ \ }} = ReLU(Conv(5ReLU(Conv({O_{WEB}})) + 4ReLU(Conv({I_L}))))
\end{array},
\end{equation}
where $5ReLU(Conv())$ and $4ReLU(Conv())$ respectively stand for five and four combinations of a convolutional layer and a ReLU. ${O_{FRB}}$ is an output of the feature refinement block, which is also an input of a reconstruction block. All mentioned convolutional kernel sizes are $3 \times 3$ and their input and output channels are 64.

\subsection{Reconstruction block}
Due to obtained low-frequency information of previous blocks, 2-layer reconstruction block can be used to deal with high-frequency information, it contains two phases. The first phase uses a sub-pixel convolutional layer to convert low-frequency information to high-frequency information. The second phase is a convolutional layer to construct a high-quality image. It can be formulated as Eq. (6). 
\begin{equation}
\begin{array}{l}
{I_L} = RB({O_{FRB}})\\
{\rm{\ \ \ \ }} = {\rm{ }}C{\rm{(}}Sub{\rm{(}}{O_{FRB}}{\rm{))}}
\end{array},
\end{equation}
where $C$ and $Sub$ denote a convolutional function and a sub-pixel convolutional function, respectively.

\begin{table}[thp]\footnotesize
\centering
\caption{Average PSNR of different methods for ×2 on Set14.}
\label{Table:1}
\addtolength{\tabcolsep}{4.8pt}
\begin{tabular}{lc}\hline
	\multirow{2}{*}{Methods}                  & Set14\\
	                                        & (PSNR) \\
	\hline
	DSRNet & 33.30\\
DSRNet without the last layer in the
  feature refinement block                                        & 33.28\\
DSRNet without the last layer and a
  residual learning operation in the feature refinement block      & 33.19\\
DSRNet without feature refinement block (threshold
  is 0.75)                                          & 33.16\\
The combination of residual enhancement
  block, complementary block and reconstruction block          & 33.06\\
The combination of residual enhancement
  block and reconstruction block                               & 33.03\\
The combination of residual enhancement
  block without last two layers, and reconstruction block      & 32.92\\
The combination of the first four layers
  in the residual enhancement block, and reconstruction block & 32.81\\
DSRNet without feature refinement block
  (threshold is 0.5)                                           & 33.10\\
DSRNet without feature refinement block
  (threshold is 0.9)                                           & 33.12\\
DSRNet without fixed dynamic gate and
  feature refinement block~                                      & 33.10\\
DSRNet without selected dynamic gate and
  feature refinement block                                    & 33.15\\
	\hline
\end{tabular}
\end{table}

\section{Experimental analysis and results}
\subsection{Datasets}
Training datasets: To keep fairness of SR methods, public dataset of DIV2K \citep{agustsson2017ntire} is used to train a DSRNet model for image super-resolution. Specifically, 800 training images of DIV2K are used as a training dataset in this paper. Each image has a LR image of $\times2$, $\times3$, $\times4$ and their corresponding HR images.

Testing datasets: To fairly test SR performance of our DSRNet, we choose popular SR datasets containing Set5 \citep{bevilacqua2012low}, Set14 \citep{zeyde2012single}, BSD100 (B100) \citep{martin2001database}, Urban100 (U100) \citep{huang2015single} and test dataset of DIV2K \citep{agustsson2017ntire} as test datasets in this paper. Also, each image in these datasets contains LR and HR images of three versions of $\times2$, $\times3$, $\times4$. In this paper, we choose Y channel of YCbCr space to conduct comparative experiments, according to LESRCNN \citep{tian2020lightweight} and CARN \citep{ahn2018fast}. 

\subsection{Experimental settings}
Training DSRNet has the following parameters. The number of steps is 6e+5. Also, original learning rate is 1e-4, it will get half every 4e+5 steps. Batch and patch sizes are set to 64, respectively. Adam is used to optimize parameters \citep{kingma2014adam}, where ${\beta _1}$ is 0.9, ${\beta _2}$ is 0.999. More parameters can be found in Ref. \citep{tian2020lightweight}. 

The DSRNet is trained on a PC with AMD EPYC 7502 32-Core Processor, 128GB RAM and four Nvidia GeForce RTX 3090, where only one GPU of 3090 is used to train our DSRNet model. Besides, all the codes run on the Ubuntu of 20.04 with Python of 3.8.13, PyTorch of 1.13.1, CUDA of 11.7 and cuDNN of 8.5.0. 

\subsection{Network analysis}
It is known that some CNNs can use deep network architectures to extract useful information to express images. However, these CNNs can share parameters to train image super-resolution models. That may suffer from challenges of unstable performance for varying scenes. To address this problem, we design a DSRNet. It depends on a residual enhancement block, wide enhancement block, feature refinement block and construction block to achieve a good performance in image super-resolution. These important components can be shown in terms of their reasonableness and validity.

Residual enhancement block: According to VGG \citep{simonyan2014very}, we can see that deep networks can facilitate more information for image applications. Inspired by that, we use stacked Conv+ReLU to extract deep information for image super-resolution. Because the first four layers in the residual enhancement block is used as a basic network, its effectiveness is not verified. Effectiveness of last two layers is proved that ‘The combination of residual enhancement block and reconstruction block’ has obtained higher PSNR than that of ‘The combination of residual enhancement block without last two layers, and reconstruction block’ in Table 1. To increase effects of different layers, we use a residual learning operation to gather outputs of the second and fourth layers to enhance memory ability of shallow layers for image super-resolution. ‘The combination of residual enhancement block without last two layers, and reconstruction block’ is higher than that of ‘The combination of the first four layers in the residual enhancement block, and reconstruction block’ in terms of PSNR, which shows effectiveness of a residual learning operation in the residual enhancement block in Table 1. 

\begin{table*}[thp]\footnotesize
\centering
\caption{PSNR and SSIM of different methods with different up-sampling factors on Set5.}
\label{Table:2}
\addtolength{\tabcolsep}{4.8pt}
\begin{tabular}{ccccc}
	\hline
 \multirow{2}{*}{Dataset} & \multirow{2}{*}{Methods} & ×2 & ×3 & ×4\\
	 & & (PSNR/SSIM)   & (PSNR/SSIM) &  (PSNR/SSIM) \\
	\hline
	  \multirow{22}{*}{Set5}   & Bicubic   & 33.66/0.9299                                  & 30.39/0.8682        & 28.42/0.8104                                                                \\
        & A+ \citep{timofte2015a+}       & 36.54/0.9544                                  & 32.58/0.9088        & 30.28/0.8603                                                                \\
        & JOR \citep{dai2015jointly}      & 36.58/0.9543                                  & 32.55/0.9067        & 30.19/0.8563                                                                \\
        & RFL \citep{schulter2015fast}      & 36.54/0.9537                                  & 32.43/0.9057        & 30.14/0.8548                                                                \\
        & SelfEx \citep{huang2015single}   & 36.49/0.9537                                  & 32.58/0.9093        & 30.31/0.8619                                                                \\
        & CSCN \citep{wang2015deep}     & 36.93/0.9552                                  & 33.10/0.9144        & 30.86/0.8732                                                                \\
        & RED \citep{mao2016image}      & 37.56/\textcolor{red}{0.9595}               & 33.70/\textcolor[rgb]{0,0.69,0.941}{0.9222}        & 31.33/0.8847                                                                \\
        & DnCNN \citep{zhang2017beyond}     & \textcolor[rgb]{0,0.69,0.941}{37.58}/\textcolor[rgb]{0,0.69,0.941}{0.9590} & \textcolor[rgb]{0,0.69,0.941}{33.75/0.9222}        & 31.40/0.8845                                                                \\
        & TNRD \citep{chen2016trainable}     & 36.86/0.9556                                  & 33.18/0.9152        & 30.85/0.8732                                                                \\
        & FDSR \citep{lu2018fast}     & 37.40/0.9513                                  & 33.68/0.9096        & 31.28/0.8658                                                                \\
        & SRCNN \citep{dong2014learning}    & 36.66/0.9542                                  & 32.75/0.9090        & 30.48/0.8628                                                                \\
        & FSRCNN \citep{dong2016accelerating}   & 37.00/0.9558                                  & 33.16/0.9140        & 30.71/0.8657                                                                \\
        & RCN \citep{shi2017structure}      & 37.17/0.9583                                  & 33.45/0.9175        & 31.11/0.8736                                                                \\
        & VDSR \citep{kim2016accurate}     & 37.53/0.9587  & 33.66/0.9213        & 31.35/0.8838                                                                \\
        & LapSRN \citep{lai2017deep}   & 37.52/\textcolor[rgb]{0,0.69,0.941}{0.9590}  & -                   & \textcolor[rgb]{0,0.69,0.941}{31.54/0.8850}                                                                \\
        & MSDEPC \citep{liu2019single}   & 37.39/0.9576                                  & 33.37/0.9184        & 31.05/0.8797                                                                \\
        & ScSR \citep{yang2010image}     & 35.78/0.9485                                  & 31.34/0.8869        & 29.07/0.8263                                                                \\
        & ESCN \citep{wang2017ensemble}     & 37.14/0.9571                                  & 33.28/0.9173        & 31.02/0.8774                                                                \\
        & DSRNet (Ours) & \textcolor{red}{37.61}/0.9584                & 
  \textcolor{red}{33.92/0.9227}
    & \textcolor{red}{31.71/0.8874} \\
	\hline
\end{tabular}
\end{table*}

\begin{table*}[!htb]\footnotesize
\centering
\caption{PSNR and SSIM of different methods with different up-sampling factors on Set14.}
\label{Table:3}
\addtolength{\tabcolsep}{4.8pt}
\begin{tabular}{ccccc}
	\hline
 \multirow{2}{*}{Dataset} & \multirow{2}{*}{Methods} & ×2 & ×3 & ×4\\
	 & & (PSNR/SSIM)   & (PSNR/SSIM) &  (PSNR/SSIM) \\
	\hline
	  \multirow{22}{*}{Set14}  & Bicubic   & 30.24/0.8688        & 27.55/0.7742                                                                & 26.00/0.7027       \\
        & A+ \citep{timofte2015a+}       & 32.28/0.9056                                                  & 29.13/0.8188                                                                & 27.32/0.7491       \\
        & JOR \citep{dai2015jointly}      & 32.38/0.9063                                                  & 29.19/0.8204                                                                & 27.27/0.7479       \\
        & RFL \citep{schulter2015fast}      & 32.26/0.9040                                                  & 29.05/0.8164                                                                & 27.24/0.7451       \\
        & SelfEx \citep{huang2015single}   & 32.22/0.9034                                                  & 29.16/0.8196                                                                & 27.40/0.7518       \\
        & CSCN \citep{wang2015deep}     & 32.56/0.9074                                                  & 29.41/0.8238                                                                & 27.64/0.7578       \\
        & RED \citep{mao2016image}      & 32.81/0.9135                                                  & 29.50/0.8334                                                                & 27.72/0.7698       \\
        & DnCNN \citep{zhang2017beyond}     & 33.03/0.9128                                                  & 29.81/0.8321                                                                & 28.04/0.7672       \\
        & TNRD \citep{chen2016trainable}     & 32.51/0.9069                                                  & 29.43/0.8232                                                                & 27.66/0.7563       \\
        & FDSR \citep{lu2018fast}     & 33.00/0.9042                                                  & 29.61/0.8179                                                                & 27.86/0.7500       \\
        & SRCNN \citep{dong2014learning}    & 32.42/0.9063                                                  & 29.28/0.8209                                                                & 27.49/0.7503       \\
        & FSRCNN \citep{dong2016accelerating}   & 32.63/0.9088                                                  & 29.43/0.8242                                                                & 27.59/0.7535       \\
        & RCN \citep{shi2017structure}      & 32.77/0.9109                                                  & 29.63/0.8269                                                                & 27.79/0.7594       \\
        & VDSR \citep{kim2016accurate}     & 33.04/0.9118                                                  & 29.76/0.8311                                                                & 28.02/0.7670       \\
        & LapSRN \citep{lai2017deep}   & 33.03/0.9124                                                  & 29.77/0.8314                                                                & 28.01/0.7674       \\
        & DRRN \citep{tai2017image}     & 33.23/0.9136                                                  & 29.96/0.8349                                                                & 28.21/0.7720       \\
        & EEDS+ \citep{wang2019end}    & 33.21/-                                & 29.85/0.8339                                                                & 28.13/0.7698       \\
        & DRFN \citep{yang2018drfn}     & \textcolor[rgb]{0,0.69,0.941}{33.29/0.9142}                  & \textcolor[rgb]{0,0.69,0.941}{30.06/0.8366}                                                                & \textcolor[rgb]{0,0.69,0.941}{28.30/0.7737}       \\
        & MSDEPC \citep{liu2019single}   & 32.94/0.9111                                                  & 29.62/0.8279                                                                & 27.79/0.7581       \\
        & ScSR \citep{yang2010image}     & 31.64/0.8940                                                  & 28.19/0.7977                                                                & 26.40/0.7218       \\
        & ESCN \citep{wang2017ensemble}     & 32.67/0.9093                                                  & 29.51/0.8264                                                                & 27.75/0.7611       \\
        & DSRNet (Ours) & \textcolor{red}{33.30/0.9145} & \textcolor{red}{30.10/0.8378}   &   \textcolor{red}{28.38/0.7760} \\
	\hline
\end{tabular}
\end{table*}

\begin{table*}[!htb]\footnotesize
\centering
\caption{PSNR and SSIM of different methods with different up-sampling factors on B100.}
\label{Table:4}
\addtolength{\tabcolsep}{4.8pt}
\begin{tabular}{ccccc}
	\hline
 \multirow{2}{*}{Dataset} & \multirow{2}{*}{Methods} & ×2 & ×3 & ×4\\
	 & & (PSNR/SSIM)   & (PSNR/SSIM) &  (PSNR/SSIM) \\
	\hline
	  \multirow{23}{*}{B100}  & Bicubic   & 29.56/0.8431                                                  & 27.21/0.7385                                                   & 25.96/0.6675                                                  \\
        & A+ \citep{timofte2015a+}       & 31.21/0.8863                                                  & 28.29/0.7835                                                   & 26.82/0.7087                                                  \\
        & JOR \citep{dai2015jointly}      & 31.22/0.8867                                                  & 28.27/0.7837                                                   & 26.79/0.7083                                                  \\
        & RFL \citep{schulter2015fast}      & 31.16/0.8840                                                  & 28.22/0.7806                                                   & 26.75/0.7054                                                  \\
        & SelfEx \citep{huang2015single}   & 31.18/0.8855                                                  & 28.29/0.7840                                                   & 26.84/0.7106                                                  \\
        & CSCN \citep{wang2015deep}     & 31.40/0.8884                                                  & 28.50/0.7885                                                   & 27.03/0.7161                                                  \\
        & DnCNN \citep{zhang2017beyond}     & 31.90/0.8961       & \textcolor[rgb]{0,0.69,0.941}{28.85/0.7981}                                                   & 27.29/0.7253                                                  \\
        & TNRD \citep{chen2016trainable}     & 31.40/0.8878                                                  & 28.50/0.7881                                                   & 27.00/0.7140                                                  \\
        & FDSR \citep{lu2018fast}     & 31.87/0.8847                                                  & 28.82/0.7797                                                   & 27.31/0.7031                                                  \\
        & SRCNN \citep{dong2014learning}    & 31.36/0.8879                                                  & 28.41/0.7863                                                   & 26.90/0.7101                                                  \\
        & FSRCNN \citep{dong2016accelerating}   & 31.53/0.8920                                                  & 28.53/0.7910                                                   & 26.98/0.7150                                                  \\
        & VDSR \citep{kim2016accurate}     & 31.90/0.8960                                                  & 28.82/0.7976                                                   & 27.29/0.7251                                                  \\
        & LapSRN \citep{lai2017deep}   & 31.80/0.8950                                                  & -                                                              & \textcolor[rgb]{0,0.69,0.941}{27.32/0.7280}                                                  \\
        & DRCN \citep{kim2016deeply}     & 31.85/0.8942                                                  & 28.80/0.7963                                                   & 27.23/0.7233                                                  \\
        & CNF \citep{ren2017image}      & \textcolor[rgb]{0,0.69,0.941}{31.91/0.8962}                                                  & 28.82/0.7980                                                   & 27.32/0.7253                                                  \\
        & MSDEPC \citep{liu2019single}   & 31.64/0.8961                                                  & 28.58/0.7981                                                   & 27.10/0.7193                                                  \\
        & ScSR \citep{yang2010image}     & 30.77/0.8744                                                  & 27.72/0.7647                                                   & 26.61/0.6983                                                  \\
        & ESCN \citep{wang2017ensemble}     & 31.54/0.8909                                                  & 28.58/0.7917                                                   & 27.11/0.7197                                                  \\
        & DSRNet (Ours) & \textcolor{red}{31.96/0.8965} & \textcolor{red}{28.90/0.8003}                   & \textcolor{red}{27.43/0.7303}                                                  \\
	\hline
\end{tabular}
\end{table*}

\begin{table*}[!htb]\footnotesize
\centering
\caption{PSNR and SSIM of different methods with different up-sampling factors on U100.}
\label{Table:5}
\addtolength{\tabcolsep}{4.8pt}
\begin{tabular}{ccccc}
	\hline
 \multirow{2}{*}{Dataset} & \multirow{2}{*}{Methods} & ×2 & ×3 & ×4\\
	 & & (PSNR/SSIM)   & (PSNR/SSIM) &  (PSNR/SSIM) \\
	\hline
	  \multirow{21}{*}{U100}  & Bicubic   & 26.88/0.8403                                    & 24.46/0.7349                                  & 23.14/0.6577                                                  \\
        & A+ \citep{timofte2015a+}       & 29.20/0.8938                                    & 26.03/0.7973                                  & 24.32/0.7183                                                  \\
        & JOR \citep{dai2015jointly}      & 29.25/0.8951                                    & 25.97/0.7972                                  & 24.29/0.7181                                                  \\
        & RFL \citep{schulter2015fast}      & 29.11/0.8904                                    & 25.86/0.7900                                  & 24.19/0.7096                                                  \\
        & SelfEx \citep{huang2015single}   & 29.54/0.8967                                    & 26.44/0.8088                                  & 24.79/0.7374                                                  \\
        & RED \citep{mao2016image}      & 30.91/0.9159                                    & 27.31/0.8303                                  & 25.35/0.7587                                                  \\
        & DnCNN \citep{zhang2017beyond}     & 30.74/0.9139                                    & 27.15/0.8276                                  & 25.20/0.7521                                                  \\
        & TNRD \citep{chen2016trainable}     & 29.70/0.8994                                    & 26.42/0.8076                                  & 24.61/0.7291                                                  \\
        & FDSR \citep{lu2018fast}     & 30.91/0.9088                                    & 27.23/0.8190                                  & 25.27/0.7417                                                  \\
        & SRCNN \citep{dong2014learning}    & 29.50/0.8946                                    & 26.24/0.7989                                  & 24.52/0.7221                                                  \\
        & FSRCNN \citep{dong2016accelerating}   & 29.88/0.9020                                    & 26.43/0.8080                                  & 24.62/0.7280                                                  \\
        & VDSR \citep{kim2016accurate}     & 30.76/0.9140                                    & 27.14/0.8279                                  & 25.18/0.7524                                                  \\
        & LapSRN \citep{lai2017deep}   & 30.41/0.9100                                    & -                                             & 25.21/0.7560                                                  \\
        & IDN \citep{hui2018fast}      & 31.27/\textcolor[rgb]{0,0.69,0.941}{0.9196  }   & 27.42/0.8359                                  & 25.41/0.7632                                                  \\
        & DRRN \citep{tai2017image}     & 31.23/0.9188                                    & 27.53/0.8378                                  & 25.44/\textcolor[rgb]{0,0.69,0.941}{0.7638}                                                  \\
        & DRFN \citep{yang2018drfn}     & 31.08/0.9179                                    & 27.43/0.8359                                  & \textcolor[rgb]{0,0.69,0.941}{25.45}/0.7629                                                  \\
        & DRCN \citep{kim2016deeply}     & 30.75/0.9133                                    & 27.15/0.8276                                  & 25.14/0.7510                                                  \\
        & MemNet \citep{tai2017memnet}   & \textcolor[rgb]{0,0.69,0.941}{31.31}/0.9195     & \textcolor[rgb]{0,0.69,0.941}{27.56/0.8376} & 25.50/0.7630                                                  \\
        & ScSR \citep{krizhevsky2017imagenet}     & 28.26/0.8828                                    & -                                             & 24.02/0.7024                                                  \\
        & DSRNet (Ours) & \textcolor{red}{31.41/\textcolor{red}{0.9209}} & \textcolor{red}{27.63/0.8402}
                              & \textcolor{red}{25.65/0.7693} \\
	\hline
\end{tabular}
\end{table*}

\begin{table*}[!htb]\footnotesize
\centering
\caption{PSNR and SSIM of different methods with different up-sampling factors on DIV2K.}
\label{Table:6}
\addtolength{\tabcolsep}{4.8pt}
\begin{tabular}{ccccc}
	\hline
 \multirow{2}{*}{Dataset} & \multirow{2}{*}{Methods} & ×2 & ×3 & ×4\\
	 & & (PSNR)   & (PSNR) &  (PSNR) \\
	\hline
	  \multirow{11}{*}{DIV2K}  & Bicubic             & 31.01       & 28.22       & 26.66       \\
        & A+ \citep{timofte2015a+}                 & 32.89       & 29.50       & 27.70       \\
        & SRCNN \citep{dong2014learning}              & 33.05       & 29.64       & 27.78       \\
        & VDSR \citep{kim2016accurate}               & 33.66       & 30.09       & 28.17       \\
        & SRResNet (L2 loss) \citep{ledig2017photo} & 34.40       & 30.82       & 28.92       \\
        & SRResNet (L1 loss) \citep{ledig2017photo} & 34.44       & 30.85       & 28.92       \\
        & EDSR \citep{lim2017enhanced}               & 35.03       & 31.26       & 29.25       \\
        & MDSR \citep{lim2017enhanced}               & 34.96       & 31.25       & 29.26       \\
        & EDSR+ \citep{lim2017enhanced}              & \textcolor[rgb]{0,0.69,0.941}{35.12}     &   \textcolor[rgb]{0,0.69,0.941}{31.39}    &   \textcolor[rgb]{0,0.69,0.941}{29.38}   \\
        & MDSR+ \citep{lim2017enhanced}              & 35.05       & 31.36       & 29.36       \\
        & DSRNet (Ours)           &   \textcolor{red}{35.67}   &  \textcolor{red}{32.04}    &   \textcolor{red}{30.15} \\
	\hline
\end{tabular}
\end{table*}
Wide enhancement block: Because CNNs may share parameters to extract features to address image applications, their performance may have challenge for varying scenes. It is known that enlarging the width can extract more complementary information to improve performance of image applications, according to GoogLeNet \citep{szegedy2015going}. Thus, we design a wide enhancement block via enlarging network width to improve effects of image super-resolution. Wide enhancement block is composed of two branches. The first branch is composed of four stacked Conv+ReLU as well as a complementary block. Its good performance is tested that ‘The combination of residual enhancement block, complementary block and reconstruction block’ has obtained higher PSNR value than that ‘The combination of residual enhancement block and reconstruction block’ in Table 1. The second branch is designed, according to the following aspects. That is, it is known that dynamic networks can learn different parameters of convolutional kernels, according to different inputs \citep{tian2023multi, zhong2022dynamic}. Inspired by that, the second branch is designed a dynamic block. The dynamic block containing two parts: a gate mechanism and dynamic gates, i.e., fixed dynamic gate and selected dynamic date. The gate mechanism is used to choose specific dynamic gate. It is composed of an average pooling, a fully connected layer, a ReLU, a fully connected layer and Softmax. Specifically, stacking an average pooling, a fully connected layer, a ReLU and a fully connected layer can be referred to Ref. \citep{zhu2020dynamic}. To normalize obtained results, a Softmax is used as the last component. Also, a Softmax is set to two results. The second result has a faster varying than that the first result in the training process. Thus, the second result is used to choose specific dynamic gate, according to Ref. \citep{zhu2020dynamic}. If the second result is lower than that of a given threshold, gate mechanism is as inputs of a fixed dynamic gate and selected dynamic gate, where fixed dynamic gate is composed of four stacked Conv+ReLU, a selected dynamic gate contains two combinations of a Conv+ReLU with size of $3 \times 3$ and Conv+ReLU with size of $1 \times 1$. Its effectiveness can be verified by ‘DSRNet without feature refinement block’ and ‘The combination of residual enhancement block, complementary block and reconstruction block’ in Table 1. It shows effectiveness of a wide enhancement block without a complementary block for image super-resolution. If the second result is higher than that of given threshold, gate mechanism is as inputs of a fixed dynamic gate. ‘DSRNet without selected dynamic gate and feature refinement block’ has obtained higher PSNR value than that ‘The combination of residual enhancement block, complementary block and reconstruction block’, which shows effectiveness of a combination of a gate mechanism and a fixed dynamic gate for image super-resolution. Gate mechanism jointed a fixed dynamic gate and selected dynamic gate is slightly superior to gate mechanism jointed a fixed dynamic gate. ‘DSRNet without feature refinement block’ and ‘DSRNet without selected dynamic gate and feature refinement block’ have near PSNR value in Table, which shows that stability of our designed dynamic block. Besides, we also show its superiority of selected dynamic gate. That is, ‘DSRNet without selected dynamic gate and feature refinement block’ has obtained an improvement of 0.04dB than that of ‘The combination of residual enhancement block, complementary block and reconstruction block’ in terms of PSNR in Table 1.  

The threshold is set to 0.75. It has the following reasons. 

1)	The second result of gate mechanism denotes more rich information than that of the first result. Thus, its threshold is set to more than 0.5. 

2)	We choose a bisection method to set the specific threshold value. We choose bisection value of 0.75 between 0.5 and 1 as the final threshold. Its superiority can be verified as follows. ‘DSRNet without feature refinement block’ of 0.75 has obtained higher PSNR value than that of ‘DSRNet without feature refinement block’ of 0.5. ‘DSRNet without feature refinement block’ of 0.75 has obtained higher PSNR value than that of ‘DSRNet without feature refinement block’ of 0.9 in Table 1.

Feature refinement block: To prevent interference of important components from wide enhancement block. We use a 6-layer stacked Conv+ReLU, according to VGG \citep{simonyan2014very}. That is verified by the following two steps. The first step uses ‘DSRNet’ and ‘DSRNet without the last layer in the feature refinement block’ to show effectiveness of the last Conv+ReLU in the feature refinement block for image super-resolution. The second step uses ‘DSRNet without the last layer and a residual learning operation in the feature refinement block’ and ‘DSRNet without feature refinement block’ with threshold of 0.75 to test effectiveness of other five Conv+ReLU for image super-resolution, where ‘DSRNet without the last layer and a residual learning operation in the feature refinement block’ has higher PSNR value than that of ‘DSRNet without feature refinement block’ as shown in Table 1. To prevent long-term dependency problem, a residual learning operation is acted between the fourth layer in the residual enhancement block and the fifth layer in the residual enhancement block. ‘DSRNet without the last layer in the feature refinement block’ has an improvement of 0.09dB than that of ‘DSRNet without the last layer and a residual learning operation in the feature refinement block’ as listed in Table 1, which shows effectiveness of a residual learning operation in the feature refinement block. 
Reconstruction block uses a sub-pixel convolution to amplify obtained features as high-frequency features. Then, a convolutional layer is used to construct a predicted high-quality image.

\begin{table*}[!htb]\footnotesize
\centering
\caption{Complexity of different SR methods for ×4 with image size of 256×256.}
\label{Table:7}
\begin{tabular}{cccc}
	\hline
	 Methods       & Parameters & Flops   & Memory   \\
	\hline
LESRCNN \citep{tian2020lightweight}  & 0.6263M    & 70.5G   & 1111MB   \\
ACNet \citep{tian2021asymmetric}    & 1.357M     & 132.82G & 1655MB   \\
CARN-M  \citep{ahn2018fast}   & 0.5335M    & 35.4G   & 838.8MB  \\
DSRNet (Ours) & 0.7461M    & 49.25G  & 710.8MB                     \\
	\hline
\end{tabular}
\end{table*}

\begin{table*}[!htb]\footnotesize
\centering
\caption{Running time (millisecond) of popular SR methods for recovering HR images with upscale factor x4 and different sizes of 256×256, 512×512 and 1024×1024.} 
\label{Table:8}
\addtolength{\tabcolsep}{4.8pt}
\begin{tabular}{ccccc}
	\hline
	   Size        & LESRCNN & ACNet & CARN-M  & DSRNet (Ours)  \\
	\hline
256 × 256   & 9.409        & 18.38      & 11.93       & 5.971          \\
512 × 512   & 40.37        & 75.79      & 45.21       & 22.96          \\
1024 × 1024 & 164.7        & -          & 195.4       & 105.3   \\
	\hline
\end{tabular}
\end{table*}

\begin{figure*}[!htb]\small
\centering
\begin{tabular}{c}
\includegraphics[scale=0.3]{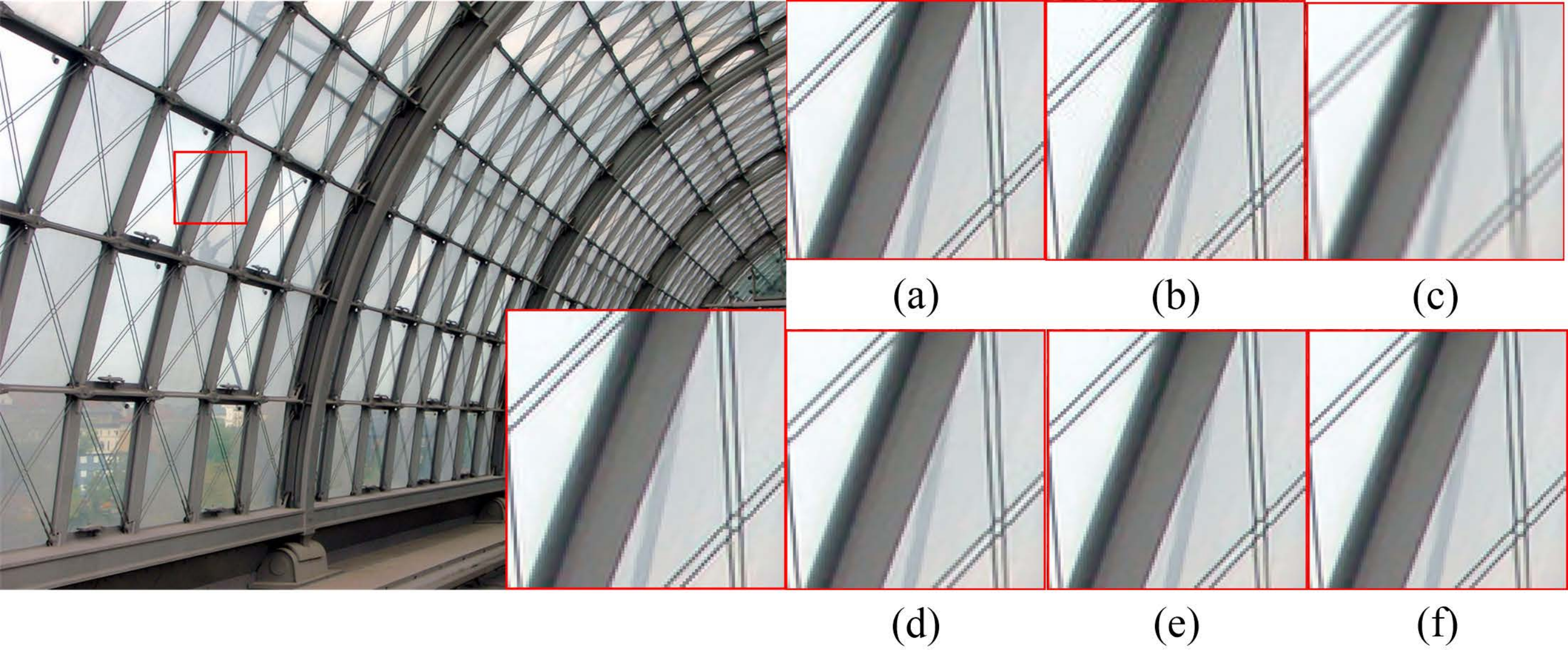}\\
\end{tabular}
\caption{Visual effects of different methods for ×2 on U100 as follows. (a) HR image, (b) Bicubic, (c) VDSR, (d) CARN-M, (e) LESRCNN and (f) DSRNet (Ours).}
\label{fig:2}
\end{figure*}

\begin{figure}[!htb]\small
\centering
\begin{tabular}{c}
\includegraphics[scale=0.5]{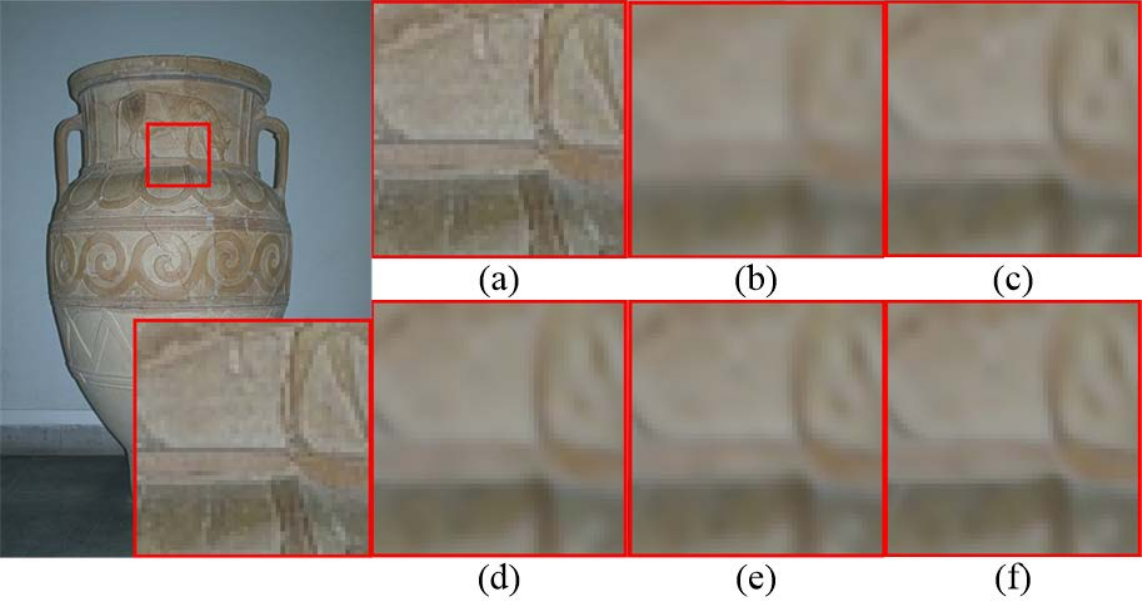}\\
\end{tabular}
\caption{Visual effects of different methods for ×4 on B100 as follows. (a) HR image, (b) Bicubic, (c) VDSR, (d) CARN-M, (e) LESRCNN and (f) DSRNet (Ours).}
\label{fig:3}
\end{figure}

\subsection{Experimental results}
To comprehensively test performance of our DSRNet for image super-resolution, we choose quantitative and qualitative analysis to evaluate its performance. In terms of quantitative analysis, we choose popular methods containing Bicubic, A+ \citep{timofte2015a+}, JOR \citep{dai2015jointly}, RFL \citep{schulter2015fast}, self-exemplars SR (SelfEx) \citep{huang2015single}, CSCN \citep{wang2015deep}, residual encoder-decoder network (RED) \citep{mao2016image}, a denoising CNN (DnCNN) \citep{zhang2017beyond}, trainable nonlinear reaction diffusion (TNRD) \citep{chen2016trainable}, fast dilated SR convolutional network (FDSR) \citep{lu2018fast}, SRCNN \citep{dong2014learning}, fast SRCNN (FSRCNN) \citep{dong2016accelerating}, residue context network (RCN) \citep{shi2017structure}, very deep SR network (VDSR) \citep{kim2016accurate}, Laplacian SR network (LapSRN) \citep{lai2017deep}, cascading residual network mobile (CARN-M) \citep{ahn2018fast}, multiscale deep encoder-decoder with phase congruency (MSDEPC) \citep{liu2019single}, lightweight enhanced SR CNN for varying scales (LESRCNN-S) \citep{tian2020lightweight}, asymmetric CNN for blind SR (ACNet-B) \citep{tian2021asymmetric}, image SR via sparse representation (ScSR) \citep{yang2010image}, ensemble based sparse coding network (ESCN) \citep{wang2017ensemble}, deep recursive residual network (DRRN) \citep{tai2017image}, end-to-end deep and shallow network (EEDS+) \citep{wang2019end}, deep recurrent fusion network (DRFN) \citep{yang2018drfn}, deeply recursive convolutional network (DRCN) \citep{kim2016deeply}, context-wise network fusion (CNF) \citep{ren2017image} information distillation network (IDN) \citep{hui2018fast}, memory network (MemNet) \citep{tai2017memnet}, SR residual network (SRResNet) \citep{ledig2017photo}, enhanced deep super-resolution network (EDSR) \citep{lim2017enhanced} and multi-scale deep super-resolution system (MDSR) \citep{lim2017enhanced} for image super-resolution as comparative methods on Set5, Set14, B100, U100 and DIV2K to test performance of our DSRNet for x2, x3 and x4. Most of these methods are lightweight. Besides, we use low-resolution images with different sizes of $256 \times 256$, $512 \times 512$ and $1024 \times 1024$ to test running time of our DSRNet for image super-resolution. To test its complexity, we choose LESRCNN, ACNet and CARN-M as comparative methods to conduct comparative experiments. As shown in Tables 2 and 3, we can see that our DSRNet almost has obtained the best or second results on x2, x3 and x4 for image super-resolution. That shows our method is very competitive in comparison with lightweight methods on small datasets for image super-resolution. As shown in Tables 4 and 5, we can see that our DSRNet almost has obtained the best or second results on x2, x3 and x4 for image super-resolution. That illustrates our method is very superior in comparison with lightweight methods on big datasets for image super-resolution. To verify its superiority, we choose high-resolution images to conduct experiments as shown in Table 6. We can see that our method has obtained the highest PSNR on DIV2K for x2, x3 and x4. Also, it has an improvement of 0.77dB than that of the second method, i.e., EDSR+ for x4. It shows that our method is very effective for recovering images of 2K. Specifically, red and blue lines are used to express the best and second results for image super-resolution as shown in Tables 2-6. Besides, we test running time and complexity of our DSRNet to verify its applicability on real digital devices, i.e., phones and cameras. Also, we can see that our DSRNet has less parameters, flops and memory in Table 7. Although it is not advantageous contrast to CARN-M in complexity, i.e., parameters and flops in Table 7, it is near one time faster and less memory than that of the CARN-M for recovering HR images of different sizes in Table 8. It is very suitable to real digital devices, i.e., phones and cameras.

Qualitative analysis is used to choose an area from a predicted HR image to amplify it as an observation area. If the observation area is clearer, its corresponding method is more effective for image super-resolution. As shown Figs. 2 and 3, we can see that our DSRNet has obtained clearer areas, which shows that our DSRNet is very competitive for visual effects. According to mentioned illustrations, we can see that our DSRNet is superior in terms of quantitative analysis of PSNR, SSIM, running time and complexity, qualitative analysis of visual effects. Besides, our method is lightweight and has fast running time for image super-resolution, which is suitable to real digital devices, i.e., phones and cameras. Although the proposed method can learn more optimal parameters, according to different scenes, it may suffer from challenges from low-resolution images with unknown destruction. We will use multi-modal technique \citep{wang2021advances} to guide a CNN for image blind super-resolution in the future.

\section{Conclusion}
In this paper, we propose a dynamic network for image super-resolution (DSRNet). That is, the first phase achieves a residual enhanced architecture to facilitate hierarchical features for image super-resolution. The second phase implements a  dynamic architecture to extract richer information to enhance applicability of obtained super-resolution model for varying application scenes. To prevent interference of components in the second phase, the third phase utilizes a stacked architecture to accurately learn obtained features, according to VGG idea. To prevent long-term dependency problem, a residual learning operation is acted on a deep layer in the third phase. To obtain predicted high-quality images, the fourth phase uses a sub-pixel convolutional layer and a single convolutional layer to construct high-quality images. Our method is lightweight and has fast running time for image super-resolution. Due to dynamic gate, the proposed network can adaptively learn more useful information for image super-resolution, according to varying scenes. We will use multi-modal technique to guide a CNN for image blind super-resolution in the future.

\section*{Acknowledgments}
This work was supported in part by National Natural Science Foundation of China under Grant 62201468, in part by the China Postdoctoral Science Foundation under Grant 2022M722599, in part by the Guangdong Basic and Applied Basic Research Foundation under Grant 2021A1515110079, in part by the Shenzhen Municipal Science and Technology Innovation Council under Grant JSGG20220831105002004, in part by Youth Science and Technology Talent Promotion Project of Jiangsu Association for Science and Technology under Grant JSTJ-2023-017, in part by the TCL Science and Technology Innovation Fund, in part by the Fundamental Research Funds for the Central Universities under Grant D5000210966.


\bibliographystyle{elsarticle-harv}
\bibliography{references}

\end{spacing}
\end{document}